\documentclass{article}
\def\e{{\rm e}}
\def\ln{{\rm ln}} 
\title{\bf 
Fat Fisher Zeroes
}
\author{ {\it W. Janke}\\
Institut f\"ur Theoretische Physik\\
Universit\"at Leipzig\\
D-04109 Leipzig\\
Germany \\
\\
{\bf and}\\
\\
{\it D.A. Johnston} and {\it M. Stathakopoulos}\\
         Dept. of Mathematics\\
         Heriot-Watt University\\
         Riccarton\\
         Edinburgh, EH14 4AS, Scotland
         }
\begin{document}
  \maketitle
%-----------------------------------------------------------------------
                      {\Large
                      \begin{abstract}
%-----------------------------------------------------------------------
%
We show that it is possible to determine
the locus of Fisher zeroes in the thermodynamic limit
for the Ising model on planar (``fat'') $\phi^4$ random graphs and
their dual quadrangulations by matching
up the real part of the high and low temperature branches of
the  expression for the free energy.  
The form of this expression for the free energy also means that
series expansion results for the zeroes may be obtained with rather less
effort than might appear necessary at first sight by simply
reverting the series expansion
of a function $g(z)$ which appears in the solution and taking a logarithm.

Unlike regular 2D lattices
where numerous unphysical critical points exist with non-standard
exponents, the Ising model on planar $\phi^4$ graphs displays only
the physical transition at $c = \exp ( - 2 \beta) = 1/4$ and
a mirror transition at $c=-1/4$ both with KPZ/DDK
exponents ($\alpha = -1 , \beta = 1/2, \gamma = 2$). 
The relation between the $\phi^4$ locus and that of the dual
quadrangulations is akin to that between the (regular) triangular and
honeycomb lattices since there is no self-duality.

%
%-----------------------------------------------------------------------
                        \end{abstract} }
%-----------------------------------------------------------------------
%
  \thispagestyle{empty}
%
%***********************************************************************
%
  \newpage
%
%-----------------------------------------------------------------------
                  \pagenumbering{arabic}
%-----------------------------------------------------------------------

\section{Introduction}

One of the more remarkable results to emerge from the study
of various statistical mechanical models coupled to 
two-dimensional quantum gravity is a solution of the Ising
model in field \cite{K,BK}. In discrete form the coupling to gravity
takes the form of the spin models living on an annealed ensemble
of triangulations or quadrangulations, or their dual planar graphs.
The partition function for the Ising model on a single graph 
$G^n$ with $n$
vertices
\begin{equation}
Z_{{\rm single}}(G^n,\beta,h) =
\sum_{\{\sigma\}} \e^{{\beta}\sum_{<i,j>} \sigma_i
\sigma_j + h \sum_i \sigma_i}\, ,
\end{equation} 
is promoted to a partition function which incorporates a sum
over some class of graphs ${\{G^n\}}$ by the coupling to gravity
\begin{equation}
Z_n(\beta,h) = \sum_{\{G^n\}} Z_{{\rm single}}(G^n,\beta,h)\, .
\end{equation} 

That introducing such an annealed sum over graphs
into the partition function should be a discrete version of 
coupling to gravity becomes clearer when one considers the
approach taken to simulating such models.
In simulations one changes both the geometry (i.e. connectivity) of the
lattice and the spins on the same timescale, so the spins affect the geometry
and vice-versa, mimicking the back-reaction of matter and gravitation in
the continuum theory.  The solution to the Ising model in \cite{K} proceeded
by first forming the grand canonical partition function
\begin{equation}
{\cal Z} = \sum_{n=1}^{\infty} \left( - 4 g c \over ( 1 - c^2)^2 \right)^n 
Z_n(\beta,h)
\label{grand}
\end{equation}   
and then noting that this could be expressed as the free energy 
\begin{equation}
{\cal Z} = - \log \int {\cal D}\phi_1~{\cal D}\phi_2~ \exp~\left( -Tr[{1\over 2}(\phi_1^2+\phi_2^2)-c\phi_1\phi_2  - \frac{g}{4}( \e^h \phi_1^4 + \e^{-h} \phi_2^4)]  \right)
\label{matint}
\end{equation}
of a matrix model, where we have written the potential that generates 
$\phi^4$ graphs. 
In the above $\phi_{1,2}$ are $N \times N$ Hermitian matrices,
$c = \exp ( - 2 \beta)$ and
the $N \to \infty$ limit is performed in order to pick out planar graphs.
The graphs of interest are generated as the Feynman diagrams of the
``action'' in equ.(\ref{matint}), which is constructed so as to weight each
edge with the correct Boltzmann weights for nearest neighbour interaction
Ising spins. Since the edges carry matrix indices the graphs in question
are ``fat'' or ribbon graphs.

The integral of equ.(\ref{matint}) can be evaluated using the results
of \cite{mehta} to give
\begin{equation}
{\cal Z} = {1\over 2} \log \left( {z\over g} \right)
-{1\over g}\int_0^z~{dt\over t}g(t)+{1\over 2g^2}\int_0^z{dt\over t}g(t)^2 ,
\label{fullpart}
\end{equation}
where $g$ is defined by
\begin{equation}
g(z)=3 c^2 z^3 +  z
\left[ \frac{1}{(1-3 z)^2}-c^2+\frac{6z(\cosh h - 1)}{(1-9 z^2)^2}
\right].
\label{geq}
\end{equation}

The implicit form of the solution may make it a little difficult
to see what is going on, but since the singularities of ${\cal Z}$ w.r.t.\ $g$
determine the asymptotics of the $Z_n$, the procedure for extracting
the thermodynamic limit is to look at the solutions to $g'(z) = 0$.
These can be explicitly determined when $h=0$ as $z_L= -1/3, \,
z_{H_{1,2}}= (1/3) [ 1 \mp \frac{1}{ \sqrt{c}} ], \,
z_{H_{3,4}}= (1/3) [ 1 \pm \frac{1}{ \sqrt{-c}} ]$ and then substituted into $g$
\begin{eqnarray}
g_L (c) &=& - \frac{1}{12} + \frac{2}{9} c^2, \nonumber \\
g_{H_{1,2}} (c) &=& \frac{2}{3} c - \frac{2}{9} c^2 \mp \frac{4} {9} \sqrt{c}~,  \\
g_{H_{3,4}} (c) &=& - \frac{2}{3} c - \frac{2}{9} c^2 \mp  \frac{4} {9} \sqrt{-c}~, \nonumber
\label{g_eq}
\end{eqnarray}
where $z_L$ is the low-temperature branch and the various $z_H$
high-temperature branches. Inserting the appropriate $g(c)$ into
the expression for $Z_n$
\begin{equation}
Z_n  \sim n^{-b} \left[{ - 4 c g (c) \over ( 1 - c^2)^2 } \right]^{-n}
\label{canon}
\end{equation}
then gives the asymptotics of the $Z_n$ and the thermodynamic
behaviour of the free energy per site $F$
\begin{equation}
F  = - \log \left[{ - 4 c g (c) \over ( 1 - c^2)^2 }\right],
\label{grand2}
\end{equation}
since if we are given a canonical partition function $Z_n$ the
associated free energy per site $F$
will be given in the thermodynamic limit by
\begin{equation}
F \sim \lim_{n \to \infty} \frac{1}{n} \log Z_n.
\label{feq}
\end{equation}  
The third order phase transition 
with the so-called KPZ/DDK \cite{KPZDDK}
exponents, $\alpha= -1, \, \beta = 1/2, \, \gamma=2$
occurs when  $g_L(c) = g_{H_{1}}(c)$
which gives a critical coupling
$c = 1/4$. It is possible to carry out a perturbative
expansion in $h$ around the $h=0$ solutions above to obtain the
magnetic critical exponents directly 
from the discretized formulation \cite{BK} and it is likewise possible
to confirm universality by solving the model on $\phi^3$ graphs.
The KPZ/DDK exponents were verified in a continuum
formalism in \cite{KPZDDK} using conformal field
theory techniques.

Given the solution of \cite{K,BK}
it is tempting to use it as a test case to investigate
various statistical mechanical ideas and methods, 
in much the same manner as the Onsager solution has served as a paradigm
over many years. One such effort was presented in \cite{jan}, where
the behaviour of the partition function zeroes for the Ising model coupled
to two-dimensional quantum gravity was investigated by series expansion
and numerical means. The study of partition function zeroes for statistical
mechanical models was initiated by Lee and Yang  for complex external fields
\cite{LYM,YL} and later extended by Fisher and others to complex temperatures
\cite{Fish}. It offers an alternative viewpoint of the approach to the thermodynamic
limit and means of extracting critical exponents.
A study of partition function zeroes
for the Ising model coupled to
two-dimensional gravity addresses several interesting questions.
It is not clear {\it a priori} that loci of partition function zeroes will
continue to lie on simple curves in the $c = \exp ( - 2 \beta )$
or $y = \exp ( -2 h)$ planes when a sum over some class of graphs, in this case
planar graphs, is folded into the partition function. Although this is
generically the case for the Onsager and related solutions on regular
two-dimensional lattices \cite{lottsashrock,lottsashrock2}, there are exceptions
such as the ``bathroom-tile'' lattice. Other sorts
of behaviour are possible too. For instance, introducing 
geometric disorder in the form of Penrose
tilings gave an complicated extended structure of temperature
zeroes away from the physical critical
point, but still gives rise to Onsager exponents \cite{prz}. Fractal lattices
on the other hand display an intricate fractal pattern of zeroes \cite{IPZ}.

The work in \cite{jan} suggested strongly that the temperature zeroes {\it did} 
lie on curves and that the field zeroes still lay on the unit circle in the 
complex $y= \exp ( - 2 h )$ plane, as in the regular lattice Onsager solution. 
Similarly, a comprehensive analytical study 
of the Lee-Yang zeroes for the 
Ising model on planar $\phi^4$ graphs was carried out in \cite{stau},
where it was found implicitly that the Lee-Yang circle theorem still
held, since the complex field singularities were 
shown to lie at purely imaginary field values.
In this paper
we concentrate on the temperature (Fisher) zeroes for the Ising model
on planar $\phi^4$ graphs and their dual quadrangulations, showing how to derive
the locus of zeroes analytically using 
the
idea that the locus
should be thought of as Stokes lines
\cite{lottsashrock,lottsashrock2,IPZ,MS,BKT}.
We compare the results with the various sorts of behaviour observed in
\cite{lottsashrock,lottsashrock2} for regular two-dimensional lattices and
also note that the form of the Ising solution means
that the various zeroes on finite planar $\phi^4$ graphs
can be extracted without evaluating a series expansion
for the full expression for ${\cal Z}$. 
In the sequel we first briefly discuss the 
general background to Lee-Yang and Fisher zeroes and the 
analytic determination of the loci of zeroes.
We then show how series expansion results
for finite graphs, such as those
in \cite{jan}, can be recovered and extended economically before we
move on to discuss obtaining the loci of
Fisher zeroes for the Ising model on $\phi^4$ graphs and their duals
analytically.
 
\section{Lee-Yang and Fisher Zeroes}

The starting point of Lee and Yang's work \cite{LYM,YL}
was the consideration
of how the non-analyticity characteristic of a phase transition
appeared from the partition function on finite
lattices or graphs, which was a polynomial
\begin{equation}
Z = \sum D_{mn} c^m y^n
\end{equation}
for a lattice with $m$ edges and $n$ vertices, again with
$c = \exp ( - 2 \beta), \, y = \exp ( -2 h )$.
They showed that the behaviour of the zeroes of this polynomial
in the complex $z$ plane,
in particular  the limiting locus as
$m,n \to \infty$,
determined the phase structure. Similarly, the behaviour of 
the zeroes in the complex $c$ plane determines the nature
of temperature driven transitions \cite{Fish}. In the latter
case, in zero external field for simplicity, the free
energy on some lattice or graph $G_n$ with $n$ nodes
and $m$ edges can be written
\begin{equation} 
F(G_n,\beta) \sim  - \ln \prod_{k=1}^m ( c - c_k (\beta)), 
\end{equation} 
which in the thermodynamic limit becomes
\begin{equation} 
F(G_{\infty},\beta) \sim - \int_L d c \rho(c) \ln (c  - c(L) ), 
\end{equation} 
where $L$ is some set of curves, or possibly regions, in the 
complex $c$ plane on which the zeroes have support and
$\rho(c)$ is the density of the zeroes there.
The singular behaviour of $\rho(c)$ as $c$
approaches the physical transition point $c_{PT}$ is related to the 
specific heat exponent $\alpha$ by
\begin{equation} 
\rho(c) \sim ( c - c_{PT})^{1 - \alpha}. 
\end{equation}

The general question of how to extract the locus of zeroes
analytically has
been considered by various authors.
It was observed in \cite{IPZ}
that such loci could be thought of as Stokes
lines separating different regions of
asymptotic behaviour of the partition
function in the complex temperature
or field planes. Across these lines
the real part of the free energy 
should be continuous and the
discontinuity in the imaginary part
should give the density of zeroes.
Shrock and collaborators \cite{lottsashrock,lottsashrock2,MS} 
have obtained  many interesting
and explicit results on the Fisher and Lee-Yang
loci for the Ising and other models on regular lattices by 
matching free energies in this manner. They also
observed that the condition 
$\Re \xi^{-1} =0$, where $\xi$ was the correlation length,
gave equivalent loci \cite{MS}.
Both this condition and the matching of free energies are 
consistent with the idea that the loci of zeroes coincide with
a change of dominant behaviour in the asymptotics.   

More recently, the case of
models with first-order transitions 
has been investigated by
Biskup {\em et al.\/} \cite{BKT} who showed rigorously  
\footnote{Under suitable technical conditions.} that
the partition function of a $d$-dimensional
statistical mechanical
model defined in a periodic volume $V = L^d$ which depends
on some complex parameter such as $c$ or $y$ can be written in terms of
complex functions $F_l$ describing $k$ different phases as
\begin{equation}
Z = \sum_l^k q_l e^{ - \beta F_l V} +  O ( e^{ - L/ L_0} e^{- \beta F V} ),
\end{equation}
where $q_l$ is the degeneracy of phase $l$, $\beta$ is the inverse temperature
and $L_0$ is of the order of the correlation length.
The various $F_l$ are the metastable
free energies per unit volume of the phases, with $\Re F_l = F$ characterising
the free energy when phase $l$ is stable.
The zeroes of the partition function are then determined
to lie within $O ( e^{ - L/ L_0 } )$ of the solutions of the
equations
\begin{eqnarray}
\Re F_{l,L}^{eff} =  \Re F_{m,L}^{eff} < \Re F_{k,L}^{eff}, \; \; \forall k \ne l,m , \nonumber \\
\beta V ( \Im  F_{l,L}  -  \Im   F_{m,L} ) = \pi \; {\rm mod} \; 2 \pi .
\label{master}
\end{eqnarray}
The equations (\ref{master}) 
are thus in perfect agreement with the idea that the loci
of zeroes should be Stokes lines, since the
zeroes of $Z$ asymptotically lie on the complex phase coexistence
curves $\Re F_{l,L} = \Re F_{m,L}$ in the complex  parameter
plane.

The specific Biskup {\em et al.\/} results apply to models with first
order transitions -- the canonical example being the field-driven transition
for the Ising model,
and we are interested here in a model with a third order
transition, so it might initially seem that
these results were inapplicable. 
We are saved by the fact that when considered
in the complex temperature plane
the transition is continuous only at the
physical point itself (and possibly some other
finite set of points). This is nicely illustrated by
looking at expressions for
the magnetization for the Ising model on the square lattice,
on fat (planar) $\phi^4$ graphs and on thin (generic) $\phi^3$ graphs: 
\begin{eqnarray}
\label{mageq}
M &=& { ( 1 + u)^{1/4} ( 1 - 6 u+ u^2 )^{1/8} \over (1 - u)^{1/2} } \; \; {\rm (square)}, \nonumber \\
M &=& { 3 ( 1 - 16 u)^{1/2} \over 3 - 8 u} \; \; {\rm (fat \; \phi^4)}, \\
M &=&   {( 1 - 3 c )^{1/2} \over (1 - 2 c  ) ( 1 + c )^{1/2}} \; \; {\rm (thin \; \phi^3)}, \nonumber
\end{eqnarray}
where $u=c^2=\exp(-4\beta)$. 
It is clear from these expressions, which apply through 
the complex extension of the low-temperature phase with $M$ zero outside,
that although $M$
will vanish continuously at the physical critical points, $u=3 -2 \sqrt{2}; \;
u= 1/16$ (i.e. $c=1/4$); $c = 1/3$ respectively
\footnote{There are further points where the magnetization
vanishes continuously: at the anti-ferromagnetic point $u=3 +2 \sqrt{2}$
and the unphysical point $u=-1$ on the square lattice; and the
unphysical point $c=-1/4$ on the planar $\phi^4$ graphs,
but these are discrete and finite in number.},
it will generally be non-zero at the phase boundary
approaching from within the low-temperature region, whereas it will be zero
approaching from outside, which is  characteristic of a first-order transition.

In summary, both general considerations about the change
of asymptotic behaviour of expansions of the partition function
in different regions of the complex temperature
or field planes 
\cite{IPZ,MS,lottsashrock,lottsashrock2} and rigorous results \cite{BKT}
lead to equ.(\ref{master}) as a 
means of determining the loci of zeroes. 

\section{Series Expansions With (two thirds) Less Pain}

To get a series expansion for ${\cal Z}$ one must in principle go back to
equ.(\ref{fullpart}) and invert (or more correctly, revert) the expression
for $g(z)$ in equ.(\ref{geq}) expanded as a series in $z$ to get an expansion $z(g)$ in
powers of $g$. This is then inserted in equ.(\ref{fullpart}) in order to obtain
the desired series from which the zeroes may be extracted. However, if one
considers the various terms in equ.(\ref{fullpart}) independently some
interesting observations can immediately be made. 
Taking each of the terms in equ.(\ref{fullpart}) separately,
\begin{eqnarray}
{\cal Z}_1: &{}&  {1\over 2}\log \left( {z\over g} \right),  \nonumber \\
{\cal Z}_2: &{}& -{1\over g}\int_0^z~{dt\over t}g(t), \\
{\cal Z}_3: &{}& {1\over 2g^2}\int_0^z{dt\over t}g(t)^2, \nonumber
\end{eqnarray}
the series expansion of each term $k=1,2,3$ can be written as 
${\cal Z}_k = \sum_{n} a^n_k A_n(c^2) g^n$ where $A_n(c^2)$ is 
{\it identical} for all the ${\cal Z}_k$. 
In addition, normalizing
$a_n = 1$ for the ${1\over 2}\log \left( {z\over g} \right) $ term 
(since any $n$ dependence in this can be put in $A_n$), one finds
\begin{eqnarray}
   a^n_1 &=& 1, \nonumber \\
   a^n_2 &=& { n \over (n+2)}, \\
   a^n_3 &=& -{ 2 n \over (n+1)}. \nonumber
\end{eqnarray}
Although the $A_n(c^2)$ which 
determine the partition function zeroes for a given
power of $g$ are the same for each ${\cal Z}_k$,
this is obscured by the different $a^n_k$ in the sum
${\cal Z}_1 + {\cal Z}_2 + {\cal Z}_3$.
 
Why should this structure be present? The solution  
given in equ.(\ref{fullpart}) comes from integrating an expression
of the form
\begin{equation}
{\cal Z} = \int_0^1 d x ( 1 - x ) \log ( f ( x )) + \ldots
\label{pre}
\end{equation}
which is common in form to all matrix models. The particular
details of a given model are encoded in the $f ( x )$
which for the Ising model with $h=0$ satisfies
\begin{equation} 
g x = \left( \frac{ 2 g f } { c} \right) \left( { 1 \over ( 1 - \frac{ 6 g f }{ c} )^2 } - c^2 \right)
+ 3 c^2 \left(\frac{ 2 g f } { c} \right)^3.
\end{equation}
The expression in equ.(\ref{fullpart}) emerges on defining $z= \frac{2 g f } { c}$
and integrating by parts.

If we expand the $\log(f( x))$ in the
integrand of equ.(\ref{pre}), then we get $\int d x (1-x)   P (g x,c) $, where 
$P (g x,c)$ is
a power-series in $g x$.
Looking at the structure of $P(g x,c)$ it is clear
that the integration over $x$ only affects
numerical factors, but {\it not} the $c^2$-polynomials which determine the zeroes.
If we now carry out  the partial integration on the
$P(g x,c)$, we obtain three
terms: $P(1,c)/2$ from the boundary corresponding to ${\cal Z}_1$
(and irrelevant additional terms); $-\int d x x  P'(g x,c)$
corresponding to ${\cal Z}_2$;  and $ \frac{1}{2}\int d x x^2  P'(g x,c)$ 
corresponding to ${\cal Z}_3$.
Carrying out the differentiation of the power series
$P'(g x ,c)$, followed by the integration above
recovers the observed values of the $a^n_k$.

The upshot of all of this is that for the purposes of extracting
partition function zeroes it is sufficient to simply consider
the expansion of $\log ( z(g)/g )$ in powers of $g$, by reverting the series
for $g(z)$,
\begin{equation}
g = (1 - c^2) z + { 6 } z^2 + {3 ( c^2 + 9 ) } z^3 + \ldots \, ,
\label{gz}
\end{equation}
to get 
\begin{equation}
\tilde z(\tilde g) = \tilde g - 6 \tilde g^2 + 3 (c^2 + 5 ) ( c^2 + 3 ) \tilde g^3 + \ldots
\label{zg}
\end{equation}
and then taking $\log ( \tilde z(\tilde g) / \tilde g )$,
where we have rescaled
$z \to (c^2-1) \tilde z$, $g \to (c^2-1)^2 \tilde g$
for algebraic convenience.  
The polynomial in $c^2$ in front of the appropriate power of $\tilde g$ will
then yield the desired Fisher zeroes. Various efficient algorithms
exist for the reversion of series (i.e. getting from equ.(\ref{gz}) to equ.(\ref{zg})
and we have used 
both the built in algorithms in Maple and
Mathematica and
one of the earliest
numerical algorithms, Newton iteration,
to revert the series for $g(z)$ \cite{GvG},
all with identical results. 

For the Newton iteration
when $h=0$ we take our starting function to be
\begin{equation}
f = (c^2-1) \tilde g (3 (c^2-1)  \tilde z-1)^2- \tilde z+(c^2  \tilde z-3 c^2 (c^2-1)^2  \tilde z^3) (3 (c^2-1)  \tilde z-1
)^2.
\end{equation}
This is then iterated with the standard Newton formula \cite{GvG}
\begin{equation}
\tilde z_{k+1} = \tilde  z_k  - \frac{f ( \tilde z_k )} {f' (\tilde z_k )}
\end{equation}
with the starting condition $\tilde z_0 = \tilde g$ (see equ.(\ref{zg}) above).
Since Newton iteration displays quadratic convergence the iteration index
$k$ is related to the order of the expansion $i$ for $\tilde z$ in $\tilde g$ by
$i = 2^k$. We thus get order $2^k - 1$ for $\tilde z ( \tilde g ) / \tilde g$
in $k$ iterations. This is both an advantage and a disadvantage
since, although 
long series are generated quite rapidly,
they are doubling in length at each iteration
which can rapidly exhaust the available memory. With the
built in functions on the other hand, 
one can proceed incrementally in the order.

To verify that $\log ( \tilde z(\tilde g)/\tilde g)$
really is sufficient to determine the zeroes correctly
we can compare 
the results for the zeroes in $c$ coming from the polynomial 
coefficient at a given
order in the expansion of $\log ( \tilde z(\tilde g)/\tilde g)$, for instance 
$\tilde g^{14}$, 
\begin{eqnarray}
&{}& {84768120 \over 7} c^{28} +5255623440 c^{26} +547079930820 c^{24} \nonumber \\
&+& 17774272305360 c^{22} +256588440930000 c^{20} +1908495144456480 c^{18} \nonumber \\
&+& 7988644803377340 c^{16} +{138652618561302240  \over 7 } c^{14} 
+30012882991193160 c^{12} \nonumber \\
&+& 28216084061998800 c^{10} +16541610886750140 c^8 +6002231595716880 c^6 \nonumber \\
&+& 1335148577661600 c^4 +173901100089600 c^2 +{ 95938227092700 \over 7},
\end{eqnarray}
with the results from the
full expression for ${\cal Z}$  from \cite{jan}. \footnote{We would like to thank
the authors of \cite{jan} for providing us with their original data.}
There is 
complete agreement 
between the numerical values of the zeroes obtained with either method
as shown in Table 1, 
\vspace{.1in}
\begin{center}
\begin{tabular}{|c|c|c|c|} \hline
$ 17.26983082 I $& $9.620359803 I$ & $4.237939134 I $      \\[.05in]
\hline
$ 3.307585457 I $ &$ 2.153341531 I$ &$ 1.952696297 I $      \\[.05in]
\hline
$ \pm .2259213695 + .3562012608 I$ & $ \pm .1989588142 + .6083974700 I$ & 
$\pm .1697220421 + .9027390050 I $   \\[.05in]
\hline
                         & $\pm .1285487771 + 1.322771774 I $&       \\[.05in]
\hline
\end{tabular}
\end{center}
\vspace{.1in}
\centerline{{\bf Table 1:} The zeroes from an expansion
of  both $\log  ( \tilde z(\tilde g)/\tilde g)$ and ${\cal Z}$ to order
$\tilde g^{14}$ are identical.}
\centerline{The complex conjugates of all 
the values shown are also zeroes.}
\vspace{.2in} 

\noindent
It is easy to obtain an expansion 
of $\log ( \tilde z(\tilde g)/\tilde g)$ in $\tilde g$ up to 
quite high order 
with relatively modest computing facilities.
The coefficient
of $\tilde g^{79}$ from such an expansion
is given in Appendix~A and we use this for comparison 
with the analytical expressions
for the loci of Fisher zeroes
in the next section.

The observations above regarding the sufficiency of 
$\log ( \tilde z(\tilde g)/\tilde g) $
for determining the
partition function zeroes also apply to both Lee-Yang zeroes and the Fisher
zeroes in non-zero field, since the general structure of the expression
for ${\cal Z} = \int_0^1 d x ( 1 - x ) \log ( f ( x )) + \ldots$
is unchanged when the field is turned on -- it is the defining equation 
for $f(x)$ which is altered. A nice confirmation of this can be obtained by
using the series expansion of $\log ( \tilde z(\tilde g) / \tilde g) $ to obtain the field zeroes in
the variable $y= \exp ( - 2 h )$ which are plotted in Fig.~1 for an
expansion up to $O(31)$
with $c=1/4$. These clearly lie on the unit circle, as they 
do for the full partition function, and are also evenly distributed.
Similarly, Fisher zeroes for the partition function in field can also be
investigated by expanding $\log ( \tilde z(\tilde g)/\tilde g)$, 
using the full expression in equ.(\ref{geq})
with $h \ne 0$. The flow observed reproduces that seen in \cite{jan} 
where the complete partition function was considered
(at much lower order).

\section{The Locus of Zeroes on $\phi^4$ Graphs}

The determination of the locus of Fisher zeroes
in the thermodynamic limit turns out to be rather straightforward,
as we now describe.
Since we wish to match $\Re F$ between the various
solution branches to obtain the loci of Fisher zeroes and
from equ.(\ref{feq}) $ F \sim \log ( g (c) )$, the equation which determines
the loci of zeroes in the thermodynamic limit is
\begin{equation} 
\log | g_L (c) | = \log | g_{H_{i}} ( c) |,
\end{equation}
or more concisely
\begin{equation}
| g_L (c) | = | g_{H_{i}} ( c) |
\label{match}
\end{equation}
where the various $g$ are given in equ.(\ref{g_eq}) and
$i=1,2,3,4$ where appropriate depending on the  region of
the complex $c$ plane. The explicit expressions arising
from substituting a complex value of $c$ into equ.(\ref{match})
are not very illuminating and we do not reproduce them here,
but they allow the resulting curves to be plotted
with Maple or Mathematica.

For comparison it is useful to refer back to the locus of Fisher
zeroes for the Ising model on a regular square lattice, which is the lima\c con
in the complex $u = c^2 = \exp ( - 4 \beta )$ plane shown in Fig.~2.
The use of $u$ has the advantage of subsuming the $c \to -c$ symmetry that 
is present in the solution and is perhaps the most natural choice of variable.
The (complex extended)
ferromagnetic phase lies inside the inner loop, the paramagnetic phase
between the loops and the antiferromagnetic phase in the exterior. The 
physical ferromagnetic and antiferromagnetic transition points lie on
the positive real axis at $u= 3 \mp \sqrt{2}$ and a multiple point with
non-standard exponents is present at $u=-1$, as already noted in the introduction
when discussing the magnetization. The lima\c con maps onto a pair of overlapping
circles in the complex $c$ plane, which is probably
a more familiar presentation.

In contrast only two phases are present in the diagram for the Ising model
on planar $\phi^4$ graphs in the $u$ plane, 
since there is no antiferromagnetic
phase in this case 
(the graphs are not loosely packed - both odd and even loops can be
present). The locus of Fisher zeroes in the
$u$ plane for the Ising model
on planar $\phi^4$ graphs is shown in Fig.~3. 
The interior of the loop is the ferromagnetic phase and the exterior the paramagnetic,
with the physical transition lying at $u=1/16$. The cusp point at
$ - \frac{1}{4}( \frac{49}{4} + 5 \sqrt{6}) = -6.124\,362 \ldots$ does
{\it not} represent an unphysical phase transition point, unlike the multiple
point on the lima\c con.
as can be confirmed by looking at the 
discriminant 
$\Delta ( g' )$ of $c g' ( z) ( 1 - z )^3 / (1 - c^2)^2$.
This will show up any non-generic points where multiple roots exist
giving phase transition points as opposed to the generic first-order boundaries
\footnote{The factor of $c/(1-c^2)^2$ has been included to match with
the factors in the free energy in equ.(\ref{grand2}) and the $(1-z)^3$ 
cancels the (irrelevant) denominator of $g'(z)$.}.
The discriminant is proportional to
\begin{equation}
\Delta ( g') \propto  { u^{7} ( 1 - 16 u)^2 \over ( 1 - u)^{16}} 
\label{disc}
\end{equation}
so the only place where non-generic behaviour appears apart
from the trivial point, $u=0$, is at 
the transition point $u = 1/16$.
At this point
the KPZ/DDK
exponents $\alpha=-1, \, \beta=1/2, \, \gamma = 2$ are manifested rather
than the flat lattice Onsager exponents. In addition to the plotted curve
a cut runs down the negative $u$ axis from the cusp to $u = - \infty$.
In this respect the locus of zeroes is more similar to that
of the Ising model on a regular triangular lattice rather than the
square lattice \cite{lottsashrock}. Obtaining zeroes for finite size
graphs from the series expansion one also finds zeroes lying along the 
negative $u$ axis.

Back in the complex $c = \exp ( - 2 \beta)$
plane one has the 
locus shown in Fig.~4 
with a physical transition point at $c = 1/4$ and a mirror image
at $c = -1/4$, both with KPZ/DDK exponents. 
We have again
omitted the cuts running up and down the imaginary axis from
the cusp points at $\pm \frac{I}{2} ( \frac{5}{2} + \sqrt{6})= \pm 
I \, 2.474\,744\,872...$ 
to $\pm I \infty$ for clarity.
In the $c$ variable
it is clear that these cuts, and the associated zeroes, appear because
of the $\sqrt{c}$ terms in the paramagnetic solutions $g_{H_i}$ in 
equ.(\ref{g_eq}). Due to
the cut on the imaginary axis
the entire left hand region of the $c$ plane
exterior to the locus
plotted in Fig.~4 forms an unphysical phase (labelled ``$O$'' in the parlance
of \cite{lottsashrock}), whereas the exterior right hand region forms
the (complex extended) paramagnetic phase. The $O$ phase vanishes
in the $u$ plane since the sections of the cut on the
imaginary axis in the $c$ plane are folded onto the negative $u$ axis. Nonetheless the presence of the cut means that it is not possible to make a circuit
of the origin in the $u$ plane in the paramagnetic region.

We have also plotted the numerically determined
roots obtained from the 
coefficient of  
$\tilde g^{79}$ in the expansion
of $\log ( \tilde z ( \tilde g ) / \tilde g)$ 
in Figs.~3,4. It is clear that the agreement with the 
analytically derived locus is very good
for the roots closest to the critical point(s). The finite-size
effects, which we have not investigated
in detail, would be interesting to explore
further since it can be seen that in the $u$ plane the roots
initially lie outside the analytic locus before moving
inside it as the cusp is approached. 

The Fisher zero locus on $\phi^4$ graphs is thus, if anything, simpler than the 
square lattice locus -- the antiferromagnetic phase is absent and there are
no unphysical transition points such as the $u=-1$ point on the square lattice.
The topology is closer to the locus of zeroes observed for
regular triangular lattices which are also not self-dual, though
the cut(s) do not extend into the ferromagnetic region 
as they do on the triangular lattice.

\section{Dual Quadrangulations}

Although the Ising model on planar $\phi^4$ graphs  does not
display an antiferromagnetic transition it does on their dual
random quadrangulations \cite{BJ}.
Again it is useful for orientational
purposes to look at the locus of zeroes 
for the square lattice Ising model which corresponds
to the case of regular quadrangulations. 
This gives the overlapping circles shown
in Fig.~5. 

The locus of zeroes for the random quadrangulations can formally
be obtained by mapping the locus of Fig.~4 in
the $c$ plane to the $c^*$ plane.
\begin{equation}
c^* = { ( 1 - c ) \over ( 1 + c ) },
\end{equation}
which may also be written as
$c^* = \tanh ( \beta)$. The identification of the phases is, however,
different for the dualized $\phi^4$ diagram and the quadrangulations.
In Fig.~6 we can see that this gives 
a ferromagnetic phase around the origin,
a crescent shaped paramagnetic phase 
and an exterior region forming
the antiferromagnetic phase.
The boundaries of the crescent
are inverse to each other in the unit circle, a section of which also
forms the image of the cut along the imaginary axis in
Fig.~4 and joins the horns of the crescent to $c =-1$ to
form the $FM/AFM$ boundary.
These horns lie at
$-0.719\,273\,115\,4... \pm I \, 0.694\,727\,418\,2...$ and just as for the 
cusp point(s) in the $u$ and $c$ planes there is
no sign of non-generic behaviour there. The numerically determined
roots have again been plotted for comparison.

One can
also obtain the 
locus for the quadrangulations 
directly by using the dualized expressions 
for $g$, replacing
$c \to ( 1 - c) / ( 1 + c)$,
\begin{eqnarray}
\tilde g_L  &=& \frac{1}{36} { 5 c^2 - 22 c + 5 \over ( 1 + c)^2}, \nonumber \\
\tilde g_{H_{1,2}} &=& - \frac{4}{9} { 2 c^2 -  c - 1 \pm  (1 + c)
\sqrt{1- c^2 }\over ( 1 + c)^2}, \\
\tilde g_{H_{3,4}} &=&  \frac{4}{9} {  c^2 +  c - 2 \pm  (1 + c)
\sqrt{c^2 -1}\over ( 1 + c)^2}, \nonumber
\end{eqnarray}
and plotting $| \tilde g_L ( c) | = | \tilde g_H ( c ) |$. 
The results are identical whichever method is used. The $\tilde g_L ( c)$
is now appropriate for the paramagnetic phase in Fig.~6, and
the $\tilde g_H ( c )$'s cover the ferromagnetic and antiferromagnetic
phases.

The dual quadrangulations display a ferromagnetic transition at
$c=3/5$ (the image of $c=1/4$) and an antiferromagnetic transition
at $c=5/3$ (the image of $c=-1/4$), both with KPZ/DDK exponents.
Looking at the dualized discriminant (c.f equ.(\ref{disc})),
\begin{equation}
\tilde \Delta  \propto   { (3 c - 5)^2 ( 5 c -3 )^2 ( 1 + c)^{14} ( 1 - c)^{14} 
\over c^{16}},
\end{equation}
we can see that this also vanishes at $c=-1$ as well as the transition
points at $3/5, 5/3$.
It would be an interesting
exercise to determine the associated exponents in full
from the matrix model formulation or series expansions,
but we do not pursue this further here.

\section{Discussion and Desiderata}

We have seen that it is possible to obtain the locus of Fisher zeroes for the Ising
model on  both planar $\phi^4$ graphs and their dual random quadrangulations 
by simply matching the moduli of the critical couplings $g(c)$ or $\tilde
g(c) $ for
the various phases. This gives the loci in Figs.~3/4 and 6 respectively which
form
an interesting contrast  to their square lattice counterparts
in Figs.~2 and 5. In particular,
the lack of self-duality on the $\phi^4$ graphs makes for less structure than 
on square lattices since there is no antiferromagnetic transition
and the general picture is more similar to
the loci seen for regular triangular lattices and the dual honeycomb
lattice \cite{lottsashrock}. 
 
Although the considerations of \cite{BKT} were for first-order transitions
the fact that the phase boundaries of the complex
extended phases generically display first-order behaviour means that
the idea of matching up the real part of different branches of the free energy
also applies to the models here.
The more general approach of \cite{IPZ,MS} 
and the idea of regarding the loci of zeroes 
as Stokes lines or complex phase boundaries
also suggests that
this will determine the loci correctly.
Comparison with series expansions results,
both from \cite{jan} and the much longer self-generated series here, 
found good agreement with the analytically determined loci.
We saw that generating such  
series expansions
to obtain either the Fisher or Lee-Yang zeroes
for the Ising model on planar graphs reduced to the reversion of a series
for $g(z)$
and taking a logarithm, 
both simple operations, and relatively long series could be produced
without undue effort.

It is perhaps worth noting that various 
properties of Fisher zero
loci discussed in \cite{lottsashrock2} 
which were derived from general considerations still apply
to the models considered here. We have:
\begin{itemize}
\item{} The loci of points where the free energy is non-analytic
is symmetric about the $\Re c$ axis 
({\it ``Theorem 1''} of  \cite{lottsashrock2}) -- applies to 
both 
$\phi^4$ graphs and quadrangulations.
\item{} If the graph has even co-ordination number
the locus of zeroes is invariant under $c \to -c$ ({\it ``Theorem 2''}
of  \cite{lottsashrock2})
-- applies to the $\phi^4$ graphs.
\item{} If the graph is bipartite the locus of zeroes is
invariant under $c \to 1/c$ ({\it ``Theorem 3''} of  \cite{lottsashrock2}) -- applies
to the random quadrangulations.
\end{itemize}
 
We have looked only cursorily at Lee-Yang zeroes in this paper, for which
a circle theorem in the complex $y = \exp ( -2 h)$ plane still appears to 
be valid
on both planar \cite{jan} and thin graphs \cite{brazil}. 
We noted that the results for the 
series expansions still hold when $h \ne 0$, and a series in $y$ 
to obtain the Lee-Yang zeroes could still
be generated by reversion of $g(z)$. 
Since the form of ${\cal Z}$ is identical
for $\phi^3$ planar graphs,
an investigation of Fisher zeroes on planar $\phi^3$ graphs and their dual
triangulations along the lines of the work here  would thus be perfectly 
feasible.
This would make for
an interesting comparison with other
regular lattices and the cases discussed here
since, as we have seen, the structure of
the loci of Fisher zeroes is highly
non-universal even when the phase transition in
the Ising model is universal.

Finally, 
we note that similar methods to those employed here may be used to obtain
the locus of Fisher zeroes for the 
(mean field) Ising and Potts model on thin random graphs
\cite{thinfish}.

\section{Acknowledgements}

W.J. and D.J. were partially supported by
ARC grant
313-ARC-XII-98/41
and  the EC IHP network
``Discrete Random Geometries: From Solid State Physics to Quantum Gravity''
{\it HPRN-CT-1999-000161}.  We would like to thank Chris Eilbeck
for useful discussions on numerical methods and Przemek Repetowicz
for his expertise on series expansions.

\newpage

\section{Appendix A}
 
The polynomial in $u$ which  appears as the coefficient of $\tilde g^{79}$
in the expansion of $\log ( \tilde z ( \tilde g ) / \tilde g)$. This was used to determine 
the zeroes plotted in Fig.~3 for comparison with the analytically
determined limiting locus.

{\tiny
\begin{eqnarray}
&-& 37139835653499716006137093004620701844717837598860560\;u^{78}\nonumber \\
&-&148105411685025580921368944483948303066640247852769586720\;u^{77}\nonumber \\
&-&219796631965060385324004998785213226455149375358517795237040\;u^{76}\nonumber \\
&-&164831966491059030819403027184073178546645417896670585118583360\;u^{75}\nonumber \\
&-&73190682361368063761602209974275459964883309418605799340499616080\;u^{74}\nonumber \\
&-&21193244550424189508255837901963813871588047644811792548846191056480\;u^{73}\nonumber \\
&-&4271156254827515627988674042672332480042287504111958952442525040283760\;u^{72}\nonumber \\
&-&627862147390520641370446437291183350790480810579190731147022437119076480\;u^{71}\nonumber \\
&-&69745140248196358151864786123993807529941628810020280214779959586962178000\;u^{70}\nonumber \\
&-&6018464875673457244143786208765208352061116125920486582535296176941628552480\;u^{69}\nonumber \\
&-&412471197424814031713421607686052888384939032979910093299169932719217502605680\;u^{68}\nonumber \\
&-&22861804785909511304850003848078876423860505468001445816621100351727090582759360\;u^{67}\nonumber \\
&-&1040388001009525627004038511792853014499415633466242701751022788738614223781869200\;u^{66}\nonumber \\
&-&39372294097894120508343007757844341394783458583098420595148520582863335631694924640\;u^{65}\nonumber \\
&-&1252669979698277617112104382833071490296757418502909447129580208467158633229292650800\;u^{64}\nonumber \\
&-&33824235941762809316986439403557899119830643621806795895083691041491223423768922205440\;u^{63}\nonumber \\
&-&781506993375339419627869035781247001736394777182618059097298519992567856933676156704880\;u^{62}\nonumber \\
&-&15562658373901209414212760616146627814737888000032257222965672288658120697186968782275040\;u^{61}\nonumber \\
&-&268811911671136955690918042236856880858371066215093407241731389769889134493580751112933200\;u^{60}\nonumber \\
&-&4050291100241419582412182577585709209008056613597679444354569945583790578947675200406223040\;u^{59}\nonumber \\
&-&53504781415966480294594562472712153356937322011357889872884583481047270602536642303088638640\;u^{58}\nonumber \\
&-&622495367645869004219933292061848993497857662753083300275019010292939581507824341349692997280\;u^{57}\nonumber \\
&-&6404543438655411633219849404298860613153085517505116213303372819518740708536174835590579287440\;u^{56}\nonumber \\
&-&58485418180717913864496799218932794580143995855302867606977886170575639976699851845262721795200\;u^{55}\nonumber \\
&-&475621336785144578141093749216959112700270248123937785852140476541108616686976720192921759659120\;u^{54}\nonumber \\
&-&3454949138847954145125771003347606489152711556862180385372143917784666252021530028539664893182560\;u^{53}\nonumber \\
&-&22479199511986118388889724740161629827026678535630939603915122091636882657168104303228159674782800\;u^{52}\nonumber \\
&-&131329868076867196195186585236624456481415618356382317417014696466013312429780652860135768267296320\;u^{51}\nonumber \\
&-&690521599841649282635736801910885058607271123389022410819911445499545307915907348287733164897752240\;u^{50}\nonumber \\
&-&3274325328024256449458979614797082828181193217762090705841294573764309335044774708753150018652173600\;u^{49}\nonumber \\
&-&14028681139432688233381147758023731741378398828505334676962034681706261933990437547881433240006857360\;u^{48}\nonumber \\
&-&54401231376357018504045387431573654084475881533965021111539531577783242990447085082088897334313973760\;u^{47}\nonumber \\
&-&191239041521657795885877250378350106358177226586893476004928070338801521234269414232669291252489401040\;u^{46}\nonumber \\
&-&610291336906248150439271697453406834444702936621298381611047533416154600700953132073526490390632405920\;u^{45}\nonumber \\
&-&1770312091536857444373426280129531056383823804457933663200346345833395486708265307228705128695195679600\;u^{44}\nonumber \\
&-&4673258693673369170099996712465402139410313870189955652194788714038526152805740081647593630983237053760\;u^{43}\nonumber \\
&-&11238340128289348531242087550331250364185389607406579052122781101900272991182180944775170566681819386000\;u^{42}\nonumber \\
&-&24643620807293562049016307383826254746709674468010311230815770648930276238060169036875457070893380721120\;u^{41}\nonumber \\
&-&49316215047681422820639798034968563472661586153859146452324880638936874014462917442830512251412077320240\;u^{40}\nonumber \\
&-&90131839322818104597333365697855009948917152222936638893065298017192961284637989321929824868625942600320\;u^{39}\nonumber \\
&-&150539307460569060574873043134235781924850278322933804772229480942138558944046423812994722573756085825360\;u^{38}\nonumber \\
&-&229903670914274936494547986274227024806559809186604659564439155029892404257079655480106334324463382885920\;u^{37}\nonumber \\
&-&321195929010989242053422405330257454914814399821523791435254317299861335551419026344418606382633018360560\;u^{36}\nonumber \\
&-&410667245136280380010581522457335103688614698402729092669357283369677965166842035371819832521182880164800\;u^{35}\nonumber \\
&-&480660884611292610900465788979182810237367769486758482580971506790193274756800002989335685501198986450960\;u^{34}\nonumber \\
&-&515127622604009076515382991320921882654805662970841545228975271887370157242400946254339190729371860734560\;u^{33}\nonumber \\
&-&505572539992143017713391549930170518971477525159833401616072175808615633494794516723314803372928619903920\;u^{32}\nonumber \\
&-&454440543980099722883378100957933438165258237407744106629390938799988591553099389645099988347996759394560\;u^{31}\nonumber \\
&-&374106736311266598660706201323205892577430894328276767390142693909422925642543062380498942843073828251600\;u^{30}\nonumber \\
&-&282037307392349122318096395401444739778301936284237145311149810212523353375933495463993869693601796997280\;u^{29}\nonumber \\
&-&194690973545520420014650890857022846138498239021284276872938877733302758744011263203822896548354353908080\;u^{28}\nonumber \\
&-&123031667663724163290143244836151984882257781474006660444783619429304123640439106921897101552503284604480\;u^{27}\nonumber \\
&-&71152689559788776900734692668351990631384530566734957851066183734605762285678691522899649638001404881040\;u^{26}\nonumber \\
&-&37645077644878972338954061259578470473781833988432377397016223534337576278963315955825141659844837247200\;u^{25}\nonumber \\
&-&18212688770152920372331067819857158399286416752811996600134670652194813938813678169137029311984688169520\;u^{24}\nonumber \\
&-&8053088045624594957085039163949872370388432536121810011655001167011932667983907251509818773391710450560\;u^{23}\nonumber \\
&-&3252470645082229356008650751768353156845940016343565742435144703608857453346650919121596251905872636880\;u^{22}\nonumber \\
&-&1199037878489113778438716271817391517512781176088717654560134880626972165041017913820459566740270632480\;u^{21}\nonumber \\
&-&403173861504807914325983089651615786813837710424803782422307586554827984880680408117617036296222317680\;u^{20}\nonumber \\
&-&123545737931826333990589263018372737196158771136705218308890409337157363897683254886286406051621360320\;u^{19}\nonumber \\
&-&34469806039566521127126491885908195430460993244436123158311930137387052183055924535551928957561135760\;u^{18}\nonumber \\
&-&8747581221336382206763336472263174629514640810049804874785909198519454278642989719373646252543135840\;u^{17}\nonumber \\
&-&2016966904073261456791906783288877971694018387744544285386966376644936811743262591541510272304280880\;u^{16}\nonumber \\
&-&422040110288334497759627356286465867221287602590081491468592540570383771671547293848313282795512320\;u^{15}\nonumber \\
&-&80036988814102662395545702571530110662277119050351283201390143378307254063442392356336074807002080\;u^{14}\nonumber \\
&-&13737160482867657888903979612359904387039625719432395863706607709157730140315331444552427440036800\;u^{13}\nonumber \\
&-&2130581489746179897063191453661629432483663971225284474690803480225230394996040706276753337037600\;u^{12}\nonumber \\
&-&298084582170998682522557843634130692118118248948169065379102074737434795961756416684382725001600\;u^{11}\nonumber \\
&-&37544966215507523322227694572841010043722835844974349906421354497206822261619859885610212620000\;u^{10}\nonumber \\
&-&4247074844108597663669728712049132720278802778418177142466258256291478726641887125832109121600\;u^{9}\nonumber \\
&-&430155384350934174771440197370772245548350430019275647467244387233120149574187173959697949600\;u^{8}\nonumber \\
&-&38846640456271627441559966602880555242219998526979599170425784619478931564006336448368211200\;u^{7}\nonumber \\
&-&3109241968691735189036674456756079403285981061864652920257364770194926524810059222607549600\;u^{6}\nonumber \\
&-&218519863192071952043562258682613228805967293880235742328799873278069522182085472663592000\;u^{5}\nonumber \\
&-&13283468415320811624119223651865154826538028015964121931579143066280124387044588883314400\;u^{4}\nonumber \\
&-&680683119521593923666628073413708550282239032204436877326746488850479986486890462153600\;u^{3}\nonumber \\
&-&28058580972988347971863829784009239496026729839677340527392289791671643157102437191200\;u^{2}\nonumber \\
&-&845523959563531370690705608441021047213998439259969130238386790640994054828513707200\;u\nonumber \\
&-&1140887404597421591776554792039966147370831359556572145202708266673791740119521072800/79 \nonumber
\end{eqnarray}
}

%-----------Figures------------------
\clearpage \newpage
\begin{figure}[htb]
\vskip 20.0truecm
\includegraphics{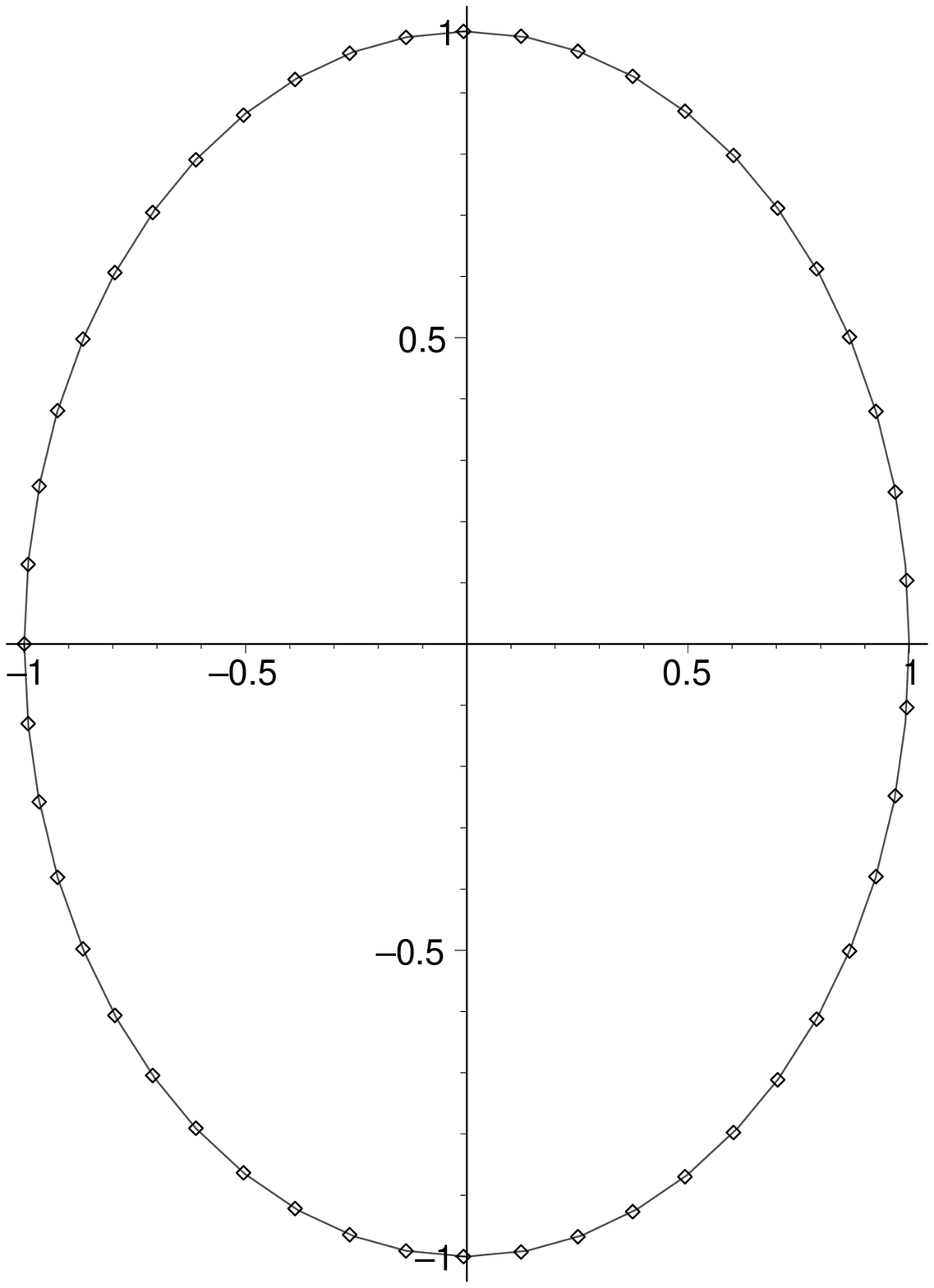}
\caption[]{\label{figm1} Lee-Yang zeroes in the complex
$y=\exp ( -2 h)$ plane for the Ising model on planar $\phi^4$ graphs
calculated from an expansion of $\tilde z(\tilde g)/\tilde g$ to order $y^{31}$. They clearly lie
on the unit circle.
}
\end{figure}
%-----------Figures------------------
%\clearpage \newpage
%
%\begin{figure}[htb]
%\vskip 20.0truecm
%\special{psfile='Cpi2pic.eps'
%   angle=0  hscale=80 vscale=60 hoffset=0  voffset= 0}
%\caption[]{\label{fig0} Fisher zeroes in field $h = i \pi / 2$
%plane for the Ising model on planar $\phi^4$ graphs,
%plotted in the $c = \exp ( - 2 \beta )$ plane. They all lie on the imaginary
%axis.
%}
%\end{figure}
%
%-----------Figures------------------    
\clearpage \newpage   
\begin{figure}[htb]
\vskip 20.0truecm
\includegraphics{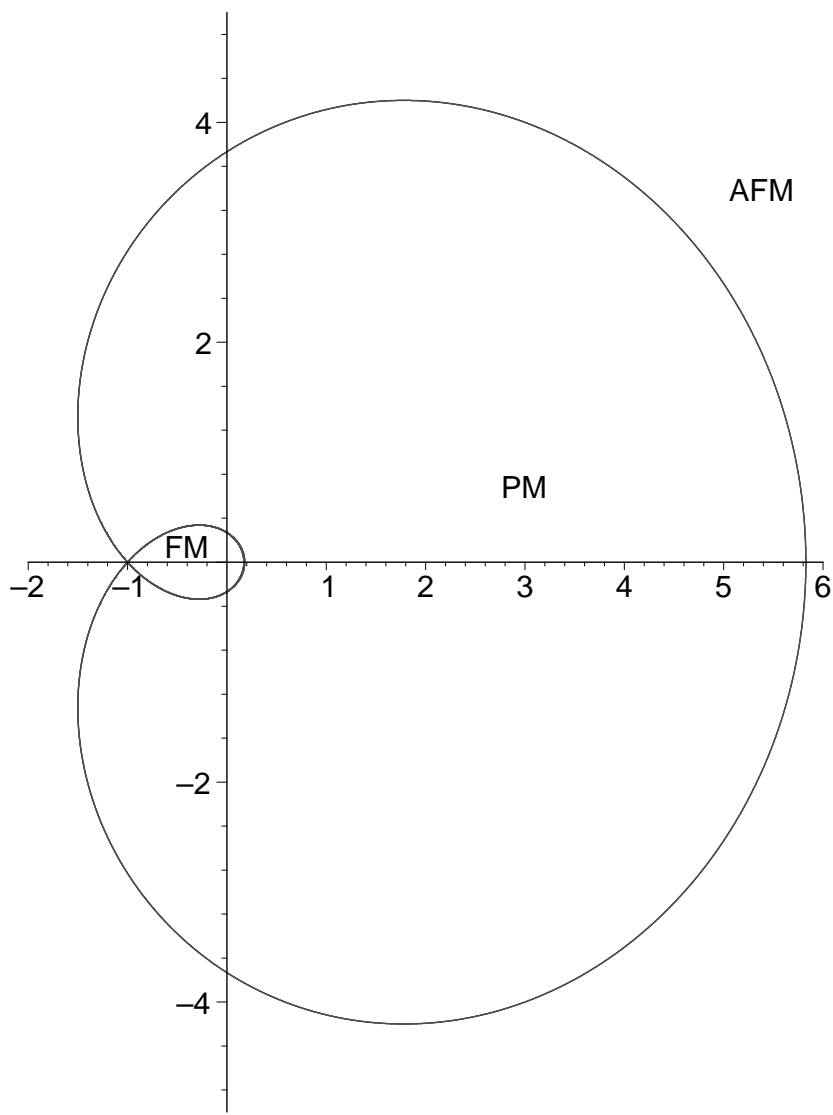}
\caption[]{\label{fig1} The locus of Fisher zeroes in the complex
$u=\exp ( -4 \beta)$ plane for the Ising model on the square lattice.
The ferromagnetic phase lies inside the inner loop, the paramagnetic phase
between the loops and the antiferromagnetic phase in the exterior.
}
\end{figure}
 
%------------------------------------ 
%------------------------------------
\clearpage \newpage
\begin{figure}[htb]
\vskip 20.0truecm
\includegraphics{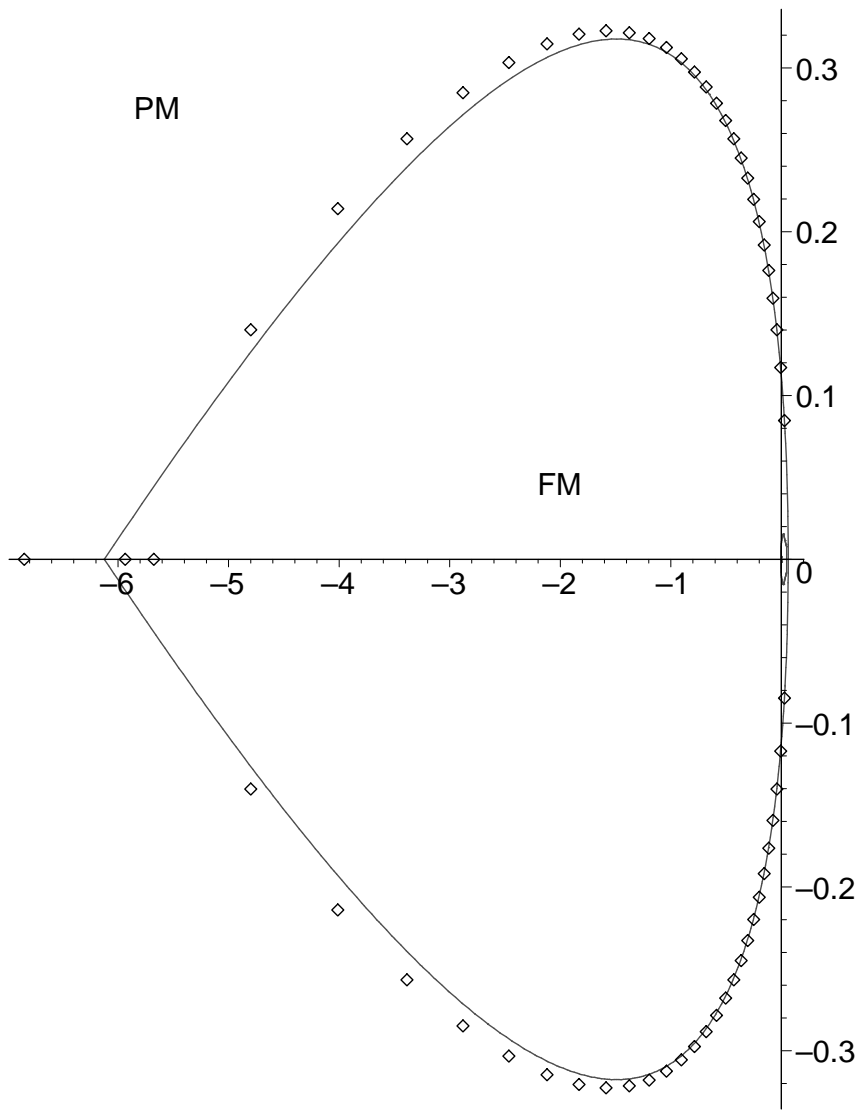}
\caption[]{\label{fig2} The locus of Fisher zeroes in the complex
$u=\exp ( -4 \beta)$ plane for the Ising model on planar $\phi^4$ graphs.
Some of the zeroes calculated from the series expansion to order $g^{79}$
in Appendix A are also plotted for comparison, though we have not plotted
the zeroes lying at large negative $u$ values on the axis.
}
\end{figure} 

%------------------------------------
\clearpage \newpage
\begin{figure}[htb]
\vskip 20.0truecm
\includegraphics{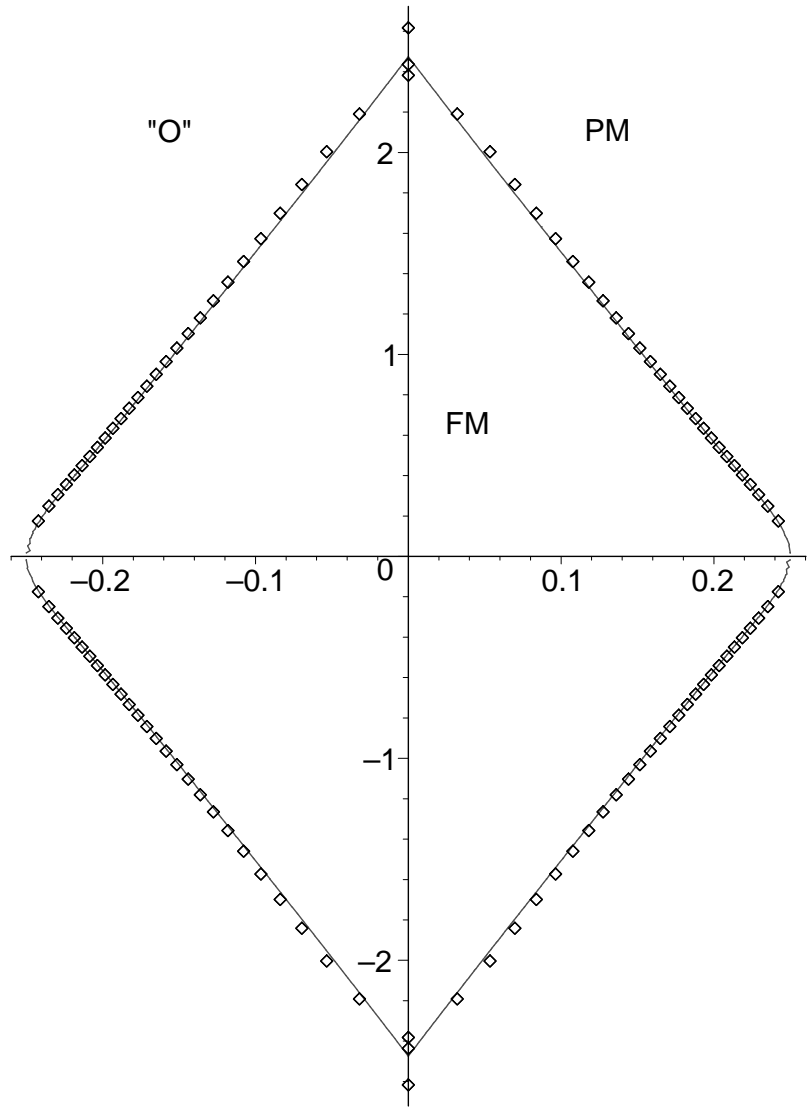}
\caption[]{\label{fig2BB} The locus of Fisher zeroes in the complex
$c=\exp ( -2 \beta)$ plane for the Ising model on planar $\phi^4$ graphs.
The interior of the curve is the ferromagnetic $FM$ region
and the exterior the paramagnetic $PM$
and unphysical $O$ phases, separated by the
cuts on the imaginary axis
which we have not plotted.
}
\end{figure}  

%------------------------------------
\clearpage \newpage
\begin{figure}[htb]
\vskip 20.0truecm
\includegraphics{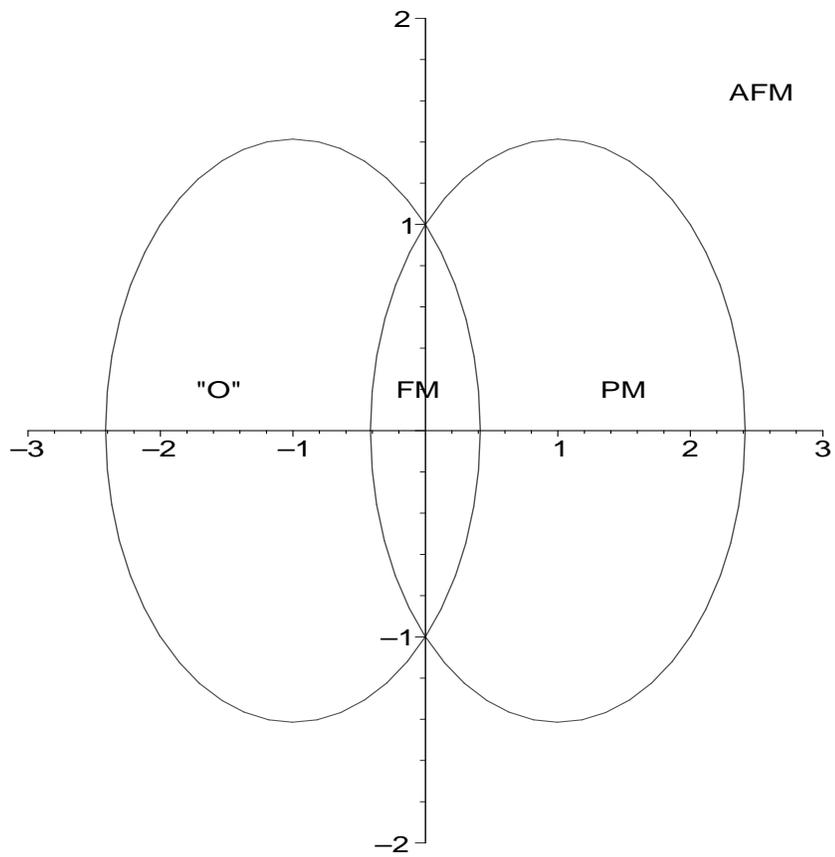}
\caption[]{\label{fig4} 
The locus of Fisher zeroes in the complex
$c$ plane for the Ising model on the square lattice. 
The $O$ region is not
connected to any physical values of $c$.
}
\end{figure}  
%------------------------------------
\clearpage \newpage
\begin{figure}[htb]
\vskip 20.0truecm
\includegraphics{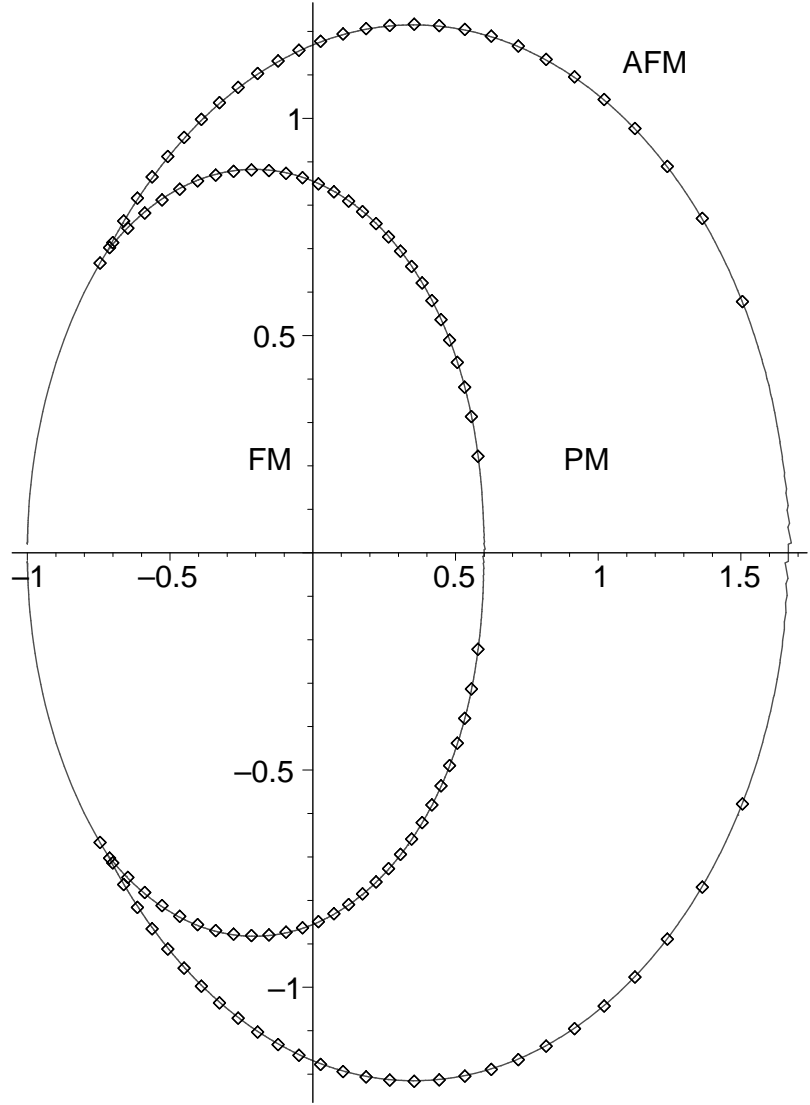}
\caption[]{\label{fig5}
The locus of Fisher zeroes in the complex
$c$ plane for the Ising model on random quadrangulations.
This is formally identical to the mapping of the locus
in Fig.~5 into the 
$c^*= ( 1 - c ) / ( 1 + c) = \tanh ( \beta)$ plane
but the identification of the phases is different. 
For quadrangulations the interior of the crescent is the paramagnetic
($PM$) phase and the two horns of the
crescent are joined to $c = -1$ by arcs of the unit circle
which are the images of the imaginary axis 
cuts in Fig.~4. These form the boundary between the inner $FM$ and 
exterior $AFM$ phases.
}
\end{figure}

\end{document}